\documentclass{article}
\usepackage{arxiv}
\usepackage[utf8]{inputenc} 
\usepackage[T1]{fontenc}    
\usepackage{url}            
\usepackage{booktabs}       
\usepackage{amsfonts}       
\usepackage{nicefrac}       
\usepackage{microtype}      
\usepackage{lipsum}		
\usepackage{graphicx}
\usepackage[sorting=none, url = false, backend=biber]{biblatex}
\usepackage{doi}
\usepackage{siunitx}
\usepackage{tikz}
\usetikzlibrary{calc}
\usetikzlibrary{intersections}
\usetikzlibrary{positioning}
\usepackage{subcaption}
\DeclareSIUnit\bar{bar}

\title{Picosecond hyperspectral fringe pattern projection for 3D surface measurement}

\author{{Sebastian Ritter} \\
	Institute of Applied Physics\\
	Friedrich Schiller University\\
	07745 Jena\\
	\texttt{sebastian.ritter@uni-jena.de} \\
	\And
	{Meritxell Cabrejo Ponce} \\
	Fraunhofer Institute of Applied Optics and Presicions Engineering IOF\\
	07745 Jena \\
	\\
	\AND
	{Nils C. Geib}\\
	Institute of Applied Physics\\
	Friedrich Schiller University\\
	07745 Jena\\
	\And 
	Sabine Häussler\\
	Fraunhofer Institute of Applied Optics and Presicions Engineering IOF\\
	07745 Jena \\
	\AND 
	Stefan Heist\\
	Fraunhofer Institute of Applied Optics and Presicions Engineering IOF\\
	07745 Jena \\
	\And 
	{Falk Eilenberger}\\
	Institute of Applied Physics\\
	Friedrich Schiller University\\
	07745 Jena\\
}

\renewcommand{\shorttitle}{\textit{arXiv} Template}

\begin{document}
\maketitle

\begin{abstract}
	Active stereovision systems for the 3D measurement of surfaces rely on the sequential projection of different fringe patterns onto the scene to robustly and accurately generate 3D surface data. This limits the temporal resolution to the time by which a sufficiently high number of patterns can be projected and recorded. By encoding patterns spectrally and recording them with a hyperspectral imager, it is possible to record several patterns in a single image, limiting the temporal resolution to only the duration of the illumination. A picosecond 3D surface measurement was demonstrated using a high pulse energy femtosecond Ti:Sa laser, spectrally broadened in a hollow core fiber, and two hyperspectral cameras recording the patterns generated by diffraction at an Echelle grating.
\end{abstract}


\section{Introduction}
Active stereovision is a well-established 3D measurement approach with the ability to measure opaque surfaces with high accuracy \cite{Geng.2011}. It has found applications in a large variety of fields in research \cite{Heist.2018}, industry \cite{Luhmann.2010}, as well as consumer electronics \cite{Zalevsky.3142006}. These measurement systems often consist of two cameras with known separation and pointing. Given the corresponding image location of identical object points in both cameras, the distance to the object point can be triangulated. To find these corresponding image points, one or more mutually independent intensity patterns are projected onto the scene. The correspondence problem can be solved by finding the maximum of the correlation function of both intensity distributions. The accuracy and robustness of the solution of the correspondence problem can be increased by sequentially projecting $N$ different patterns onto the same scene, where ideally $N \geq 10$ \cite{Psarakis.2005, Lutzke.2013}. The sequence is usually a temporal series of pattern projection and frame recording. This temporal multiplexing limits the time resolution of the measurement systems to the speed at which the different patterns can be projected and recorded, e.g.\ the frame rate of the camera. Commonly, fringe patterns are generated using a projector with a rotating wheel in conjunction with high-speed cameras \cite{Heist.2016,Hyun.2018,Zhang.2019}. Using this method, temporal resolutions in the range of a hundred microseconds have been demonstrated \cite{Heist.2018}. 

By color-coding the patterns with a single spectrally multiplexed pattern, the temporal resolution is limited only by the active time of the illumination. Hence, a temporal resolution on ultrafast timescales can be achieved by using a femtosecond laser as a light source. 

\begin{figure}
    \centering
    \includegraphics[scale = 0.55]{ 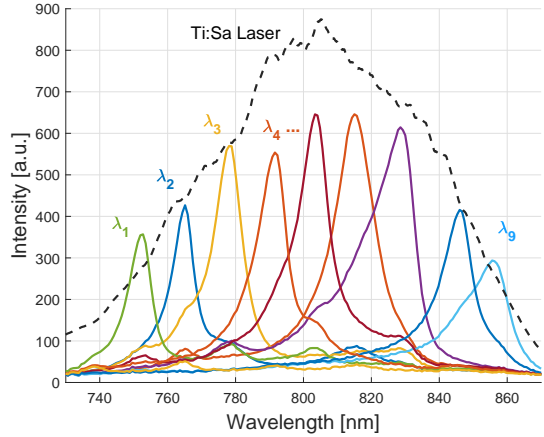}
    \caption{Signal intensity in the channels $\lambda_{1...9}$ of a hyperspectral camera for excitation with a modelocked Ti:Sa Laser. Measured by illuminating the camera with narrow sections of the laser spectrum, which is shown as the dashed line.}
    \label{fig:ximea_measured}
\end{figure}

For such an ultrafast spectrally multiplexed pattern generation, several key requirements must be met. To differentiate between spectrally multiplexed patterns, the spatial distribution as well as the spectral characteristics of the intensity patterns reflected from the scene have to be evaluated. Hyperspectral cameras offer this functionality. The working principle of these cameras is similar to a conventional RGB camera. In the case of RGB cameras, the color information is generated using spectral filters on the camera sensor arranged in a 2-by-2 Bayer matrix, from which a full scale color image is reconstructed \cite{Bayer.351975, PjotrStoevelaar.2020}. This approach is enhanced for hyperspectral imaging by increasing the size of the filter matrix, e.g.\ to 25 sub-pixels with different spectral filters. By using Fabry-Perot interferometric filters, each color channel can be engineered to measure a narrow wavelength range \cite{Tack.3202019}. The spectral sensitivity of a suitable hyperspectral camera that is illuminated with a Ti:Sa laser in the near infrared region is shown in Fig. \ref{fig:ximea_measured}. Hyperspectral cameras have found use in diverse research fields, for example medicine \cite{Zhang.2020, Lu.2014}, archaeology \cite{Liang.2012}, botany \cite{ZHANG.2018}, as well as applications in defense \cite{Chen.2014}, art conservation \cite{Cucci.2016}, and waste management \cite{GiuseppeBonifazi.2014}. As the different spectral channels are located on the same imaging sensor, no complex matching between the channels has to be done.

To utilize the ultrashort pulse duration of a femotsecond laser for 3D, the necessary measurement data has to be generated in a single laser pulse. This requires, depending on camera sensitivity and the size of the illuminated object, a pulse energy in the range of hundreds of microjoules. 

The number of patterns that can be projected is limited by the spectral bandwidth of the light source. Modelocked Ti:Sa lasers provide ultrashort pulses with a conversely broad spectral bandwidth \cite{Stingl.1995}, both of which are desired in this application. Considering the spectral sensitivities of the different color channels in the hyperspectral camera, a bandwidth of at around \SI{100}{\nano\meter} is required to illuminate $N=10$ channels. This can be achieved by spectrally broadening the laser pulses, as in our case the initial bandwidth is insufficient. For this, self phase modulation (SPM) can be utilized. Confinement of the high intensity laser pulses in a hollow core fiber enhances this effect \cite{JanRothhardt.2011, SteffenHadrich.2012}. If a noble gas is used as the nonlinear medium, the ultrashort duration and high energy of the pulses is maintained. This method is well established in the generation of high energy few-fs laser pulses \cite{Bohle.2014, Jacqmin.2015}.

By combining the spectral resolution of the hyperspectral cameras with the broad spectrum and high energy of a nonlinearily broadened femtosecond laser, a single shot 3D measurement with a temporal resolution below \SI{100}{\nano\second} was demonstrated. The quality of the projected fringes was evaluated based on the 3D reconstruction a surface moving at high subsonic speeds.

\section{Methods}
\subsection{Spectrally broadened light source}
A modelocked Ti:Sa oscillator, amplified using chirped pulse amplification, fulfills the requirements best, given the output of spectrally broad femtosecond pulses with energies in the range of millijoules in the near-infrared. In this case, a commercial chirped pulse amplification (CPA) system was used (Coherent Legend), generating \SI{2}{\milli\joule} sub-ps pulses with \SI{60}{\nano\meter} bandwidth at \SI{800}{\nano\meter} center wavelength.

Given the spectral sensitivity of the cameras shown in Fig.  \ref{fig:ximea_measured}, approximately seven channels are illuminated using this laser. By spectrally broadening the ultrashort pulses, a larger number of channels can be illuminated to increase the accuracy and robustness of the surface measurement. 

\begin{figure}[h]
    \centering
    \begin{subfigure}[t]{0.42\textwidth}
        \centering
        \begin{tikzpicture}[scale = 0.45]
            \coordinate (center) at (4,4);
\coordinate (left) at (2.5,4);
\coordinate (left2) at (2.5,0.5);
\coordinate (right) at (5.5,4);
\coordinate (right2) at (5.5,0.5);

\draw [fill = cyan!25, thick]
    (center) circle (3);

\path [fill = white]
    (center) circle (1.5);
    
\node (air) at (center) {\scriptsize Air};

\node (glass) at ($(center) - (0,2.25)$) {\scriptsize Glass};

\draw [dashed] (left) -- (left2);

\draw [dashed] (right) -- (right2);

\draw [<->] (left2) -- (right2) node [pos = 0.5, anchor = north] {\scriptsize  \SI{200}{\micro\meter}};
            \vspace{1cm}
        \end{tikzpicture}
        \caption{Typical layout and size of a hollow core fiber. The light is guided in the air filled core in the center of the fiber.}
        \label{fig:hcf_schema}
    \end{subfigure}
    \hfill
    \begin{subfigure}[t]{0.42\textwidth}
       \centering
        \includegraphics[scale = 0.375]{ 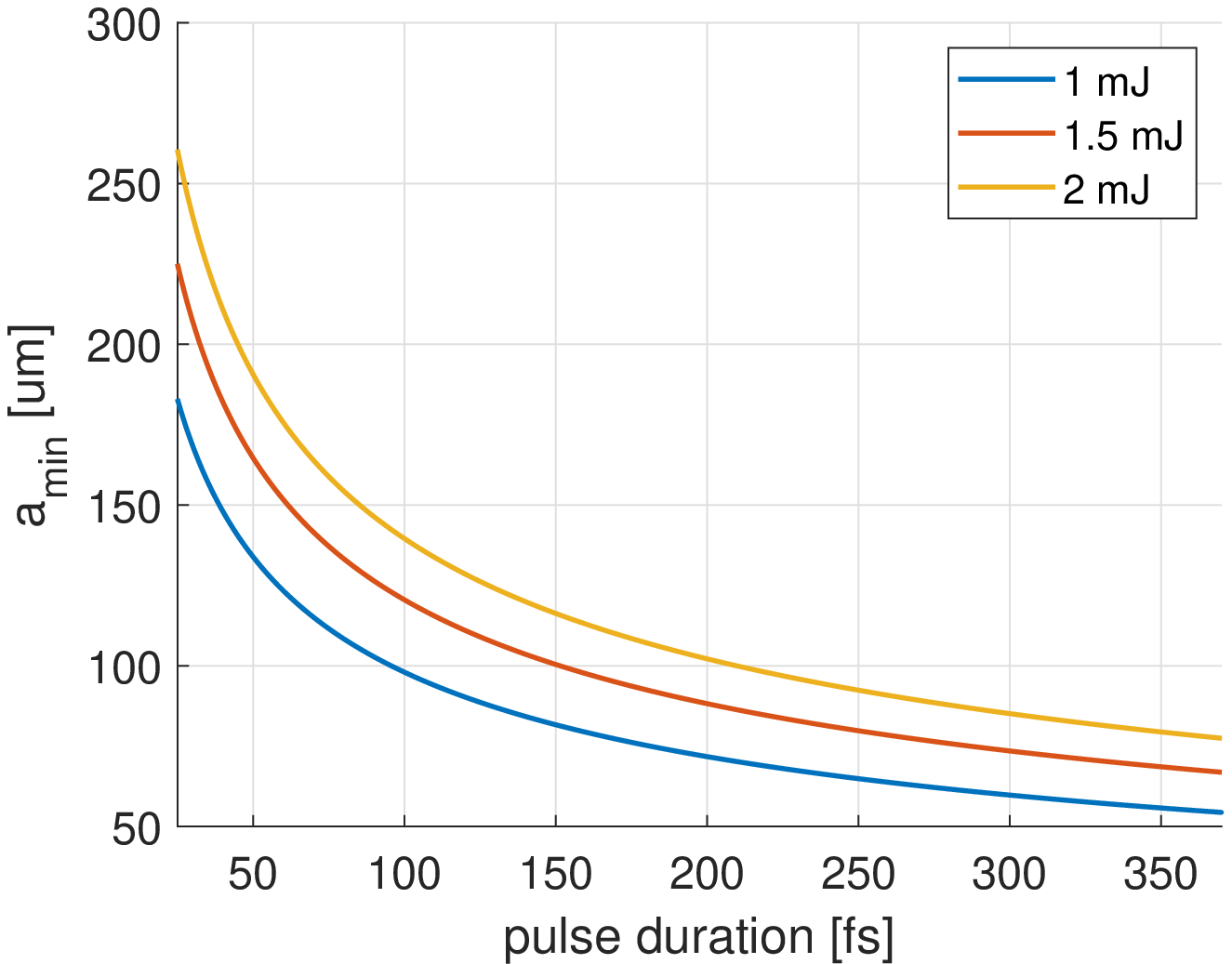}
        \caption{Lower limit of the core radius given by ionization effects for different pulse duration $T_0$ and energies $E_0$ in an Argon filled fiber.}
        \label{fig:ionization_graph} 
    \end{subfigure}
    \hfill
    \caption{Structure and characteristics of the hollow core fiber used for the spectral broadening of ultrashort pulses with high energy.}
    \label{fig:hcf_theory}
\end{figure}

Spectral broadening of high energy ultrashort pulses has been demonstrated in noble gas filled hollow core fibers, like the one shown in Fig. \ref{fig:hcf_schema}. The pulses are confined in the core of the fiber by Fresnel reflection off the inner core wall at grazing incidence, resulting in Bessel like modes. The absorption coefficient $\alpha$ for propagation in the hollow core depends on the radius of the core $a$ and the ratio between the refractive indices of cladding and core $\nu$ \cite{Marcatili.1964}
\begin{equation}
    \alpha = \left( \frac{2.405}{2\pi} \right)^2 \frac{\lambda ^2}{a^3} \frac{1}{2} \frac{(\nu^2 + 1)}{\sqrt{\nu^2-1}}.
    \label{eqn:alpha}
\end{equation}

As the pulses are not propagating in glass, high powers can be guided without destroying the fiber. In case of the fiber setup used for this application, the pulse evolution is mainly governed by SPM stemming from the nonlinear Kerr index of the noble gas.

The nonlinear effect of SPM on the spectrum of a ultrashort pulse is well described by the nonlinear Schrödinger equation \cite{Agrawal.2013}. The nonlinear phase acquired during propagation of a pulse introduces a change in the instantaneous frequency, thus spectrally broadening the pulse. For this application, the pulse energy and duration is not affected significantly. This spectral broadening is limited by, among other effects, parasitic self-focusing during propagation stemming from the SPM, as well as the ionization of the nonlinear medium. Quantitatively, these two limitations manifest in a maximal pressure of the nonlinear medium and a minimal radius of the fiber core respectively \cite{Vozzi.2005}. The maximum pressure $p_{\mathrm{max}}$
\begin{equation}
    p_{\mathrm{max}} = 0.15 \frac{\lambda_0^2}{\nu_2 P_0}
    \label{eqn:max_pressure}
\end{equation}
depends on the center wavelength $\lambda_0$ and peak power $P_0$ of the pulses, as well as the pressure dependent nonlinear Kerr index $\nu_2$. As the nonlinear Kerr index increases with gas pressure, higher pressures lead to stronger spectral broadening. 
Numerical models for the ionization caused by ultrashort electrical fields \cite{Ammosov.1992} provide an upper limit on the gas pressure based on the  duration $T_0$ and energy $E_0$ of the pulse. The effect of the gaseous nonlinear medium is expressed in the constant $A_{\mathrm{gas}}$. The exponents $\beta, \gamma$ are fitted to $0.51$ and $0.45$ respectively, so that the minimal radius $a_{\mathrm{min}}$ of the fiber core can be determined by

\begin{equation}
    a_{\mathrm{min}} = A_{\mathrm{gas}} T_0^{-\gamma}E_0^\beta .
    \label{eqn:ion}
\end{equation}
This is illustrated in Fig. \ref{fig:ionization_graph} for varying pulse duration and typical energies.
A smaller radius leads to a reduced effective mode area and a stronger field confinement and thus to increased broadening. These two parameters have to be balanced to achieve an optimal and stable broadening . 

\subsection{Fringe pattern generation}
For the generation of sufficiently variant and dense stripe patterns in each of the spectral channels of a hyperspectral camera an Echelle grating was chosen. These low groove density blazed gratings can be engineered to exhibit a higher efficiency for high number diffraction orders \cite{Pedrotti.2008}. They are commonly used in high resolution spectrometers \cite{Keliher.1976, Barnard.1993}. The high order diffraction maxima are spectrally dependent and thus generate a hyperspectral fringe pattern.

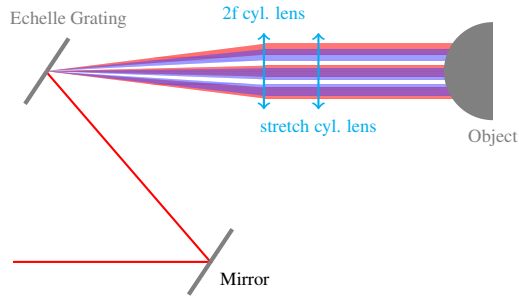
\begin{figure}[h]
    \centering
    \begin{tikzpicture}[scale = 1.45]
       
\coordinate (fiber4) at (4.2,1);
\coordinate (grating1) at (6,1);
\coordinate (lens1) at (5.75,2);
\coordinate (lens2) at (6.25,2);
\coordinate (mask1) at (5.75,2.75);
\coordinate (mask2) at (6.25,2.75);
\coordinate (lens3) at (5.75,3.5);
\coordinate (lens4) at (6.25,3.5);
\coordinate (grating2) at (6,4.5);

\coordinate (grating3) at (4.5,2.75);
\coordinate (lens5) at (6.5, 2.5);
\coordinate (lens6) at (6.5,3);
\coordinate (screen1) at (8.5, 2);
\coordinate (screen11) at (8.5, 2.5);
\coordinate (screen2) at (8.5, 3.5);
\coordinate (screen21) at (8.5, 3);
\coordinate (screen3) at (8.6, 2.9);
\coordinate (lens51) at (6.5, 2.6);
\coordinate (lens52) at (6.5, 2.7);
\coordinate (lens53) at (6.5, 2.8);
\coordinate (lens54) at (6.5, 2.9);

\coordinate (s51) at (8.5, 2.6);
\coordinate (s52) at (8.5, 2.7);
\coordinate (s53) at (8.5, 2.8);
\coordinate (s54) at (8.5, 2.9);

\coordinate (g1) at (8.75, 2.15);

\draw[gray, ultra thick, opacity = 1] 
    (grating1) 
    node [anchor = north west, color = black, opacity = 1] {\scriptsize Mirror};

\draw[red, thick]
    (fiber4) -- ($(grating1) + (0.01,0)$) -- (grating3);
    
\shade[top  color = red, bottom color = red, opacity = 0.5]
    (grating3) -- (lens5) -- (lens51) -- cycle;
    
\shade[top color = blue!70, bottom color = blue!70, opacity = 0.5]
    (grating3) -- ($(lens5) + (0, 0.025)$) -- ($(lens51) + (0, 0.025)$) -- cycle;
    
\shade[top  color = red, bottom color = red, opacity = 0.5]
    (grating3) -- (lens52) -- (lens53) -- cycle;
    
\shade[top color = blue!70, bottom color = blue!70, opacity = 0.5]
    (grating3) -- ($(lens52) - (0, 0.025)$) -- ($(lens53) - (0, 0.025)$) -- cycle;

\shade[top  color = red, bottom color = red, opacity = 0.5]
    (grating3) -- (lens54) -- (lens6) -- cycle;

\shade[top color = blue!70, bottom color = blue!70, opacity = 0.5]
    (grating3) -- ($(lens54) - (0, 0.05)$) -- ($(lens6) - (0, 0.05)$) -- cycle;

\shade[top  color = red, bottom color = red, opacity = 0.5]
   (screen11) -- (lens5) -- (lens51) -- (s51) -- cycle; 

\shade[top color = blue!70, bottom color = blue!70, opacity = 0.5]
    ($(screen11)  + (0, 0.025)$) -- ($(lens5) + (0, 0.025)$) -- ($(lens51) + (0, 0.025)$) -- ($(s51)  + (0, 0.025)$) -- cycle;

\shade[top  color = red, bottom color = red, opacity = 0.5]
   (s52) -- (lens52) -- (lens53) -- (s53) -- cycle; 

\shade[top color = blue!70, bottom color = blue!70, opacity = 0.5]
    ($(s52)  - (0, 0.025)$) -- ($(lens52) - (0, 0.025)$) -- ($(lens53) - (0, 0.025)$) -- ($(s53)  - (0, 0.025)$) -- cycle;

\shade[top  color = red, bottom color = red, opacity = 0.5]
   (s54) -- (lens54) -- (lens6) -- (screen21) -- cycle; 

\shade[top color = blue!70, bottom color = blue!70, opacity = 0.5]
   ($(s54) - (0, 0.05)$) -- ($(lens54) - (0, 0.05)$) -- ($(lens6) - (0, 0.05)$) -- ($(screen21)  - (0, 0.05)$)-- cycle;

\fill[color = gray] ($(screen21) + (0.1, 0.2)$) arc(90:270:0.45) node [anchor = north] {\scriptsize Object} -- cycle;

    
\draw[gray, ultra thick, opacity = 1] 
    ($(grating3) + (0.01,0)$)-- +(-0.2, -0.3);
\draw[gray, ultra thick, opacity = 1] 
    ($(grating3) + (0.01,0)$) -- +(0.2, 0.3)
    node [anchor = south] {\scriptsize Echelle Grating};

\draw[gray, ultra thick, opacity = 1] 
    ($(grating1) + (0.01,0)$)-- +(-0.2, -0.3);
\draw[gray, ultra thick, opacity = 1] 
    ($(grating1) + (0.01,0)$) -- +(0.2, 0.3);

\draw[<->, cyan, thick]
    ($(lens5) + (0,-0.1)$) -- ($(lens6) + (0,0.1)$)
    node [anchor = south] {\scriptsize 2f cyl. lens};
    
\draw[<->, cyan, thick]
    ($(lens5) + (0.5,-0.1)$)  node [anchor = north] {\scriptsize stretch cyl. lens} -- ($(lens6) + (0.5,0.1)$)
   ;
    
       \vspace{1cm}
    \end{tikzpicture}
    \caption{Fringe pattern generation by high order diffraction from an Echelle grating.}
    \label{fig:diff_pattern}
\end{figure}

To project the interference fringes onto a scene, a 2f-setup consisting of a cylindrical lens is used for the horizontal direction along the diffraction. By using a perpendicular cylindrical lens, the pattern is stretched  vertically to cover a larger measurement area. In Fig.  \ref{fig:diff_pattern}, this is illustrated for three diffraction maxima and two wavelengths. 

\subsection{Active stereovision 3D surface measurement}
In general, a 3D measurement setup aims to generate depth values $z$ as a function of the position $(x,y)$, which can be expressed as a matrix $z_{ij} = (x_i, y_j)$ that represents the measured surface in 3D space. Furthermore, if each point on the surface has a scalar value $f$ associated to it, one can define a point cloud $P_i = (x_i, y_i, z_i, f_i)$, where $f$ can for example represent the measured intensity or the surface reflectance at the $i$th surface point \cite{Geng.2011}. This can be extended to vectorial quantities $\vec{v_i} = (\lambda_i^1,\lambda_i^2,\lambda_i^3,...,\lambda_i^n)$ of dimension $n$, e.g. if the spectral characteristics are measured for $n$ different wavelengths $\lambda$ \cite{Heist.2018}. 

The approach of active stereovision aims to generate the $z$ values by triangulation. A scene is recorded by two cameras with known separation, pointing and imaging behavior. For the triangulation, the position of the corresponding image points of each object point have to be known, commonly referred in literature as the correspondence problem in computer vision \cite{Martin.1988}. It can be solved by finding the maximum of the cross correlation of the two images \cite{Psarakis.2005}. By artificially generating an intensity distribution in the scene by projecting a fringe pattern, the solution of the correspondence is aided by giving each object point a unique brightness sequence. This is in our case a unique spectral distribution \cite{Geng.2011}. With this spectral multiplexing a large number of patterns can be projected onto the scene at the same time, so that for each given wavelength region, a different pattern is projected.

The two camera images are rectified by rotation and translation of the coordinate system \cite{Hartley.2003} to reduce the dimensionality of the correspondence problem. The cameras are modeled by the geometric imaging of pinhole cameras, which requires calibration. A detailed description of the reconstruction algorithm can be found in  \cite{Heist.2018}.



To evaluate the suitability of the fringe patterns measured in the different spectral channels, the 3D reconstruction was calculated for objects with known geometry, i.e. a diffusely reflective plane. The temporal resolution was tested by measuring the plane object moving at high speed. A suitable parameter for this is the standard deviation of the measured position from the expected plane, as well as the percentage of measurement values that can be attributed to reconstruction artifacts or outliers. The expected plane is fitted using a linear regression. Reconstruction artifacts are defined as measurement points with a distance from the fitted plane that is larger than \SI{5}{\milli\meter}, to exclude measurement values that are within the uncertainty of the measured plane.  

\section{Experimental Work}
\subsection{Light source}
The output of the commerical CPA system was coupled into a hollow core fiber of \SI{50}{\centi\meter} length with an inner core radius of \SI{220}{\micro\meter}, which is in the vicinity of the critical radius defined by equation \ref{eqn:ion}. To fill the fiber core with the nonlinear medium, it was placed in a vacuum chamber with two Brewster windows for entry and exit ports. By repeatedly evacuating the chamber and filling it with gas, a high purity of the nonlinear medium was ensured. For the experiment, Argon was chosen as the nonlinear medium. It provides a sufficiently high nonlinear Kerr index, reasonable maximum pressure of \SI{835}{\milli\bar}, and is readily available. For the given fiber parameters a transmission over \SI{90}{\percent} in the linear regime is possible, based on equation \ref{eqn:alpha}.

\begin{figure}[h]
    \centering
    \begin{subfigure}[t]{0.42\textwidth}
        \centering
        \includegraphics[scale = 0.375]{ 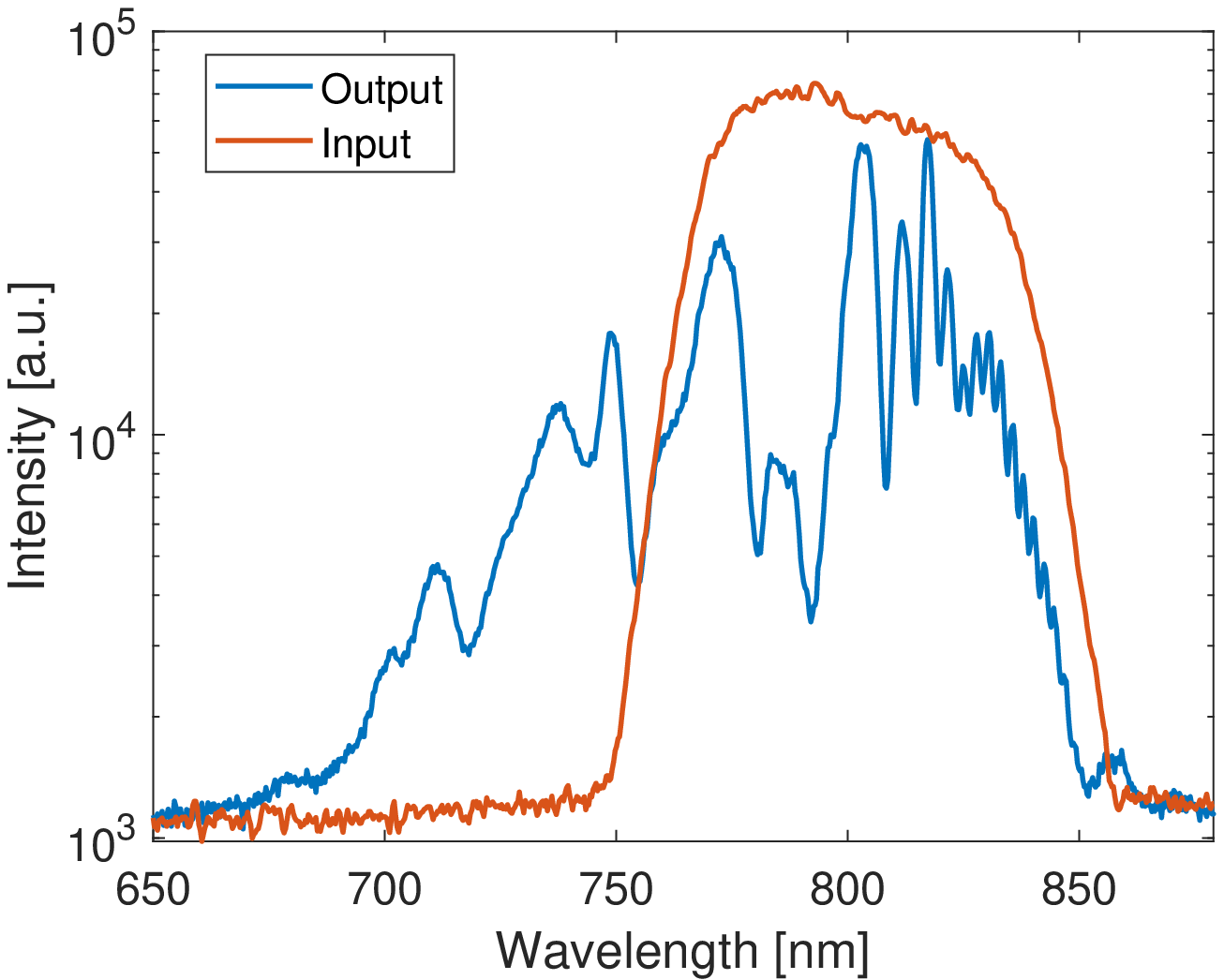}
        \caption{Optimized spectrum with a usable spectrum spanning from \SI{700}{\nano\meter} to nearly \SI{850}{\nano\meter}.}
        \label{fig:mj_final}
    \end{subfigure}
    \hfill
    \begin{subfigure}[t]{0.42\textwidth}
       \centering
        \includegraphics[scale = 0.375]{ 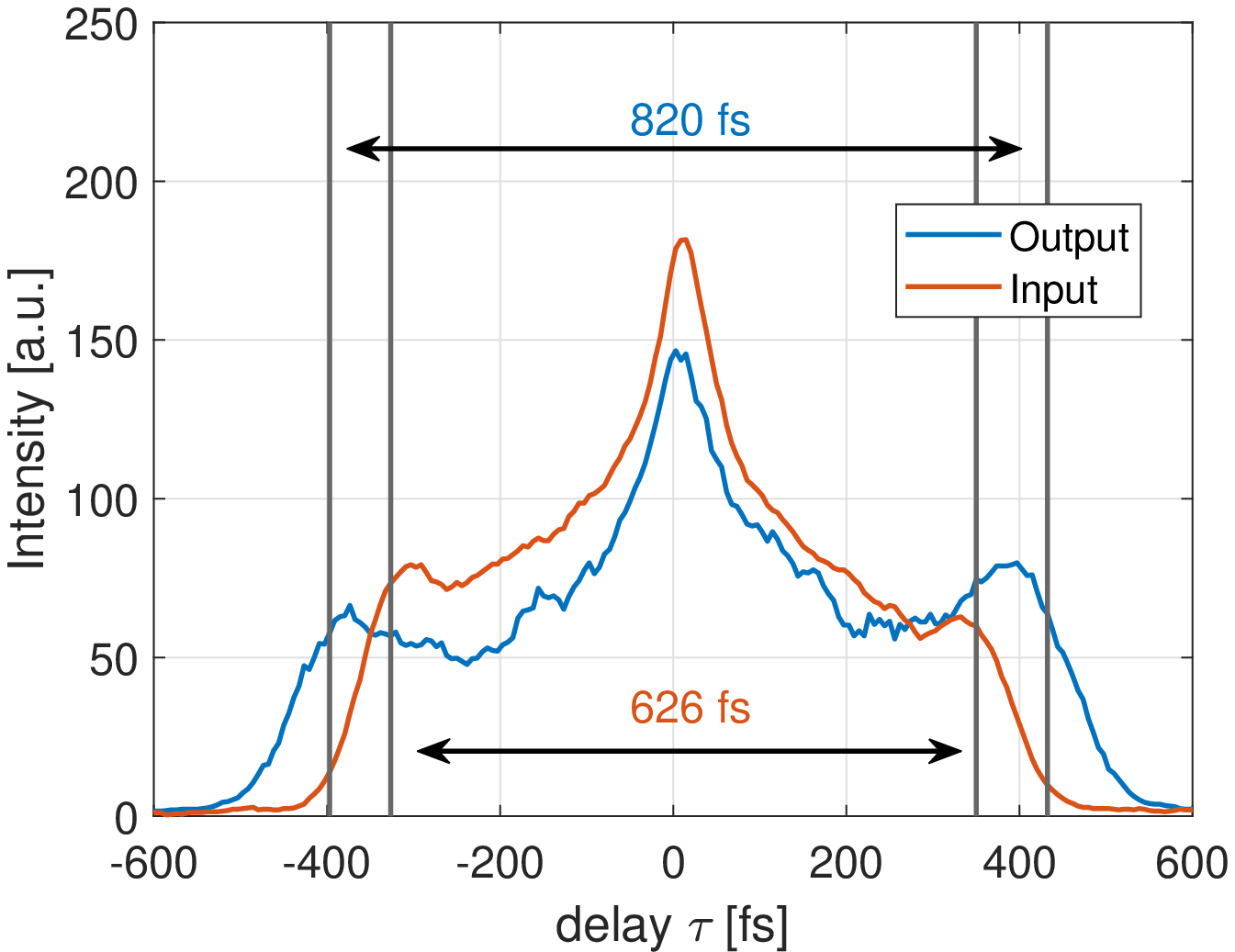}
        \caption{Temporal Pulse Profile before and after spectral broadening, measured using autocorrelation.}
        \label{fig:res_fiber_AC} 
    \end{subfigure}
    \hfill
    \caption{Output of the fiber broadening}
    \label{fig:output_fiber}
\end{figure}

The output of the hollow core fiber was evaluated using a spectrometer (OceanOptics USB2000+), a camera and a autocorrelation setup (APE PulseCheck). 
The transmission of the broadening setup was measured with a thermal powermeter as approximately \SI{50}{\percent}. In Fig. \ref{fig:output_fiber}, the measurement of the spectrum and the pulse duration is shown. The spectral width is increased to nearly \SI{150}{\nano\meter} of usable spectrum, while the output temporal profile only shows a marginally increased pulse duration. It has to be noted that the input pulse is intentionally stretched, which was used to balance the spectral broadening with the parasitic effect of self-focusing reducing transmission. Pulse duration before and after the fiber broadening is below \SI{1}{\pico\second}.

\subsection{Pattern generation}

\begin{figure}[h]
    \centering
    \begin{subfigure}[t]{0.49\textwidth}
        \includegraphics[scale = 0.42]{ 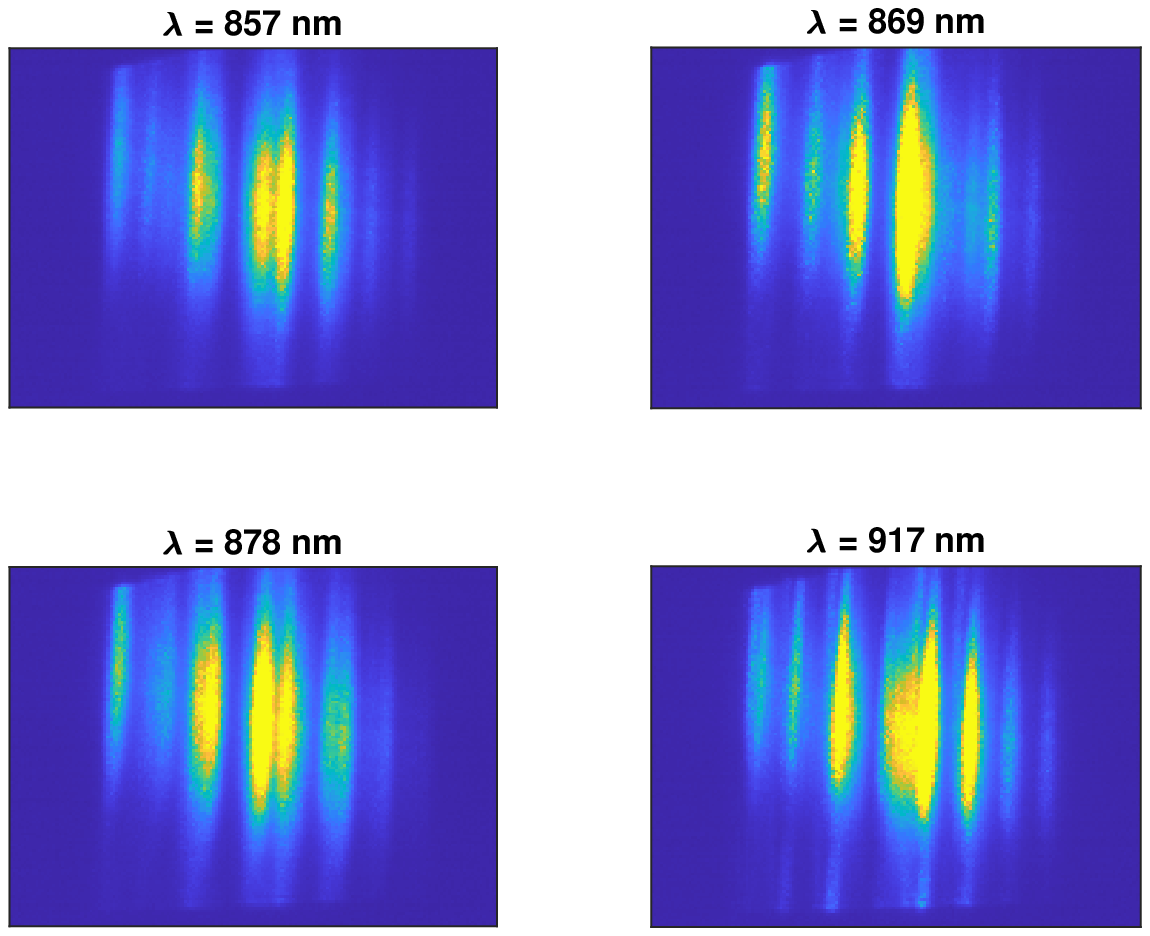}
        \caption{View in camera 0}
    \end{subfigure}
    \hfill
    \begin{subfigure}[t]{0.49\textwidth}
        \includegraphics[scale = 0.42]{ 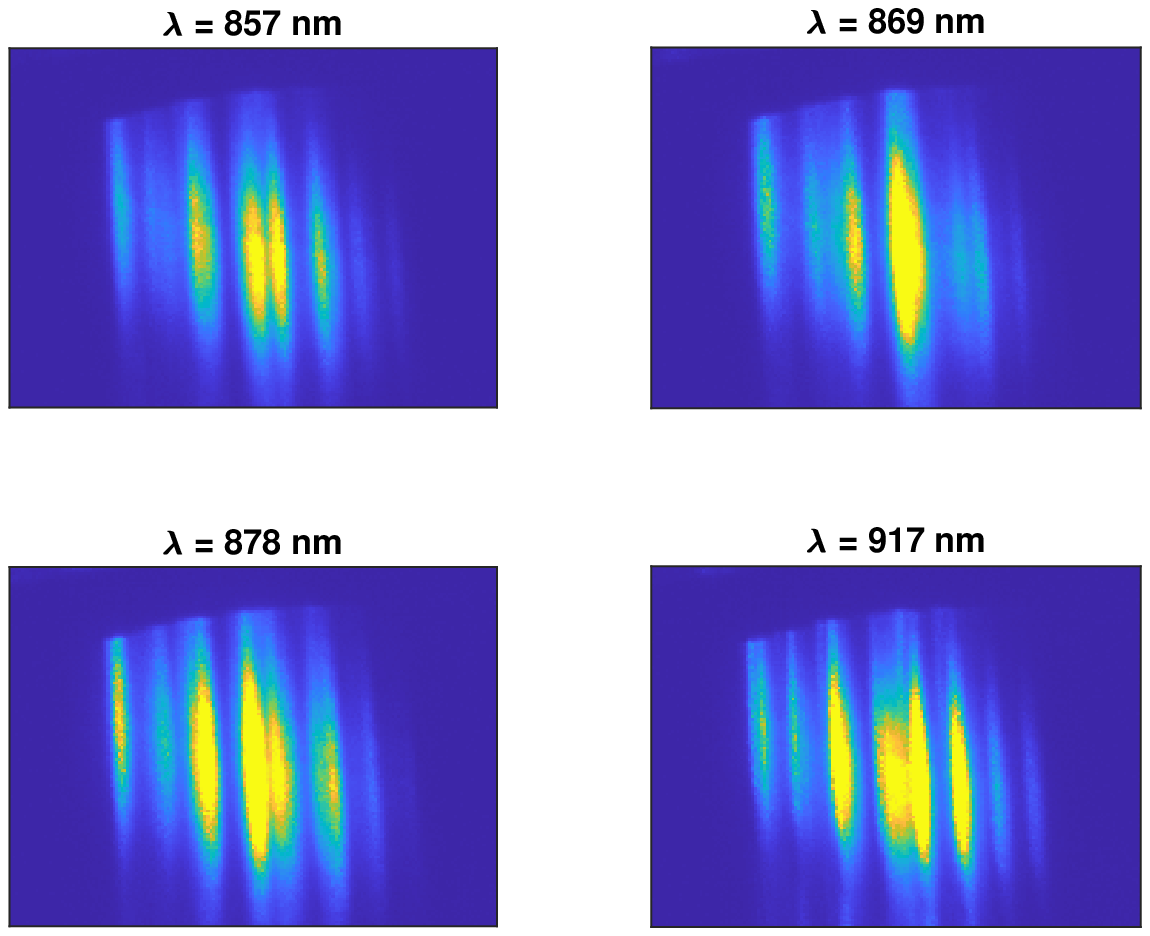}
        \caption{View in camera 1}
    \end{subfigure}
    
    \caption{Examplary fringe patterns generated with overlapping high order diffraction maxima of a Echelle grating in the four spectral channels with the highest intensity for both cameras. }
    \label{fig:patterns_grating}
\end{figure}

\begin{figure}[h]
    \centering
    \includegraphics[scale = 0.42, trim=0 0 0 0, clip]{ 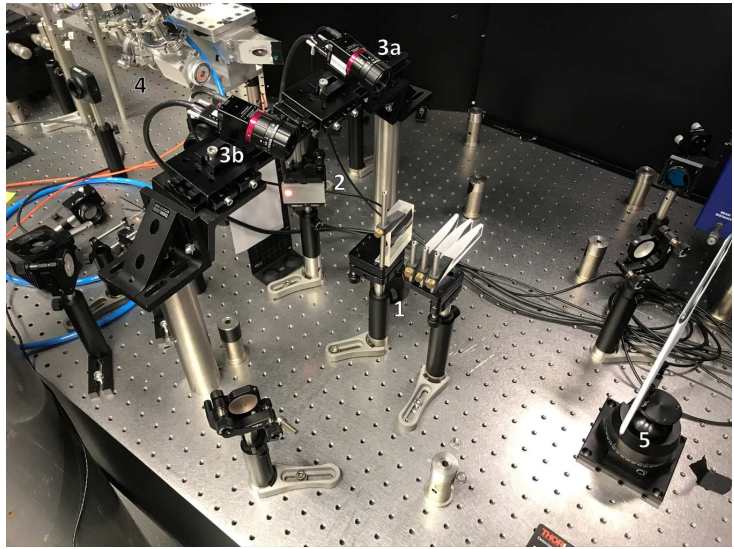}
    \caption{Experimental setup for the generation of ultrafast broadband fringe patterns. Top left: Vacuum chamber (4) with fiber for spectral broadening. Middle: Echelle grating (2) and projection lenses (1). Above: Hyperspectral cameras (3a, 3b) aimed at target (5) in the right of the image.}
    \label{fig:irl_setup}
\end{figure}

An Echelle grating (Thorlabs GE2550-0363) with a groove density of \SI{31.6}{\per\meter} and a blaze angle of \SI{63}{\degree} is used in perpendicular incidence of the laser beam. The so-generated high order diffraction fringes are projected as described above onto a plane and diffusely reflecting surface and were recorded with the hyperspectral cameras (Ximea MQ022HG) with a 2048-by-1088 subpixel resolution.  The complete setup of the diffractive pattern generation is shown in Fig. \ref{fig:irl_setup}. The Echelle grating is visible as a gray rectangle in the middle of the image. In Fig. \ref{fig:patterns_grating} the four spectral channels with the overall highest intensity are shown as measured by both cameras.

Due to the non-uniform spectral and spatial profile of the laser pulse, the fringes show an intensity variance in each channel, as well as between the channels. The spectral dependence of the hyperspectral camera, given by the different transmission of the Fabry-Perot etalons and the varying efficiency of a silicon based sensor in the near-einfrared spectral region, also contributes to this. 

\subsection{3D reconstruction}

\begin{figure}[h]
    \hspace{\fill}
    \begin{subfigure}{0.4 \linewidth}
        \centering
        \includegraphics[trim = {20 70 20 0}, clip, scale = 0.38]{ 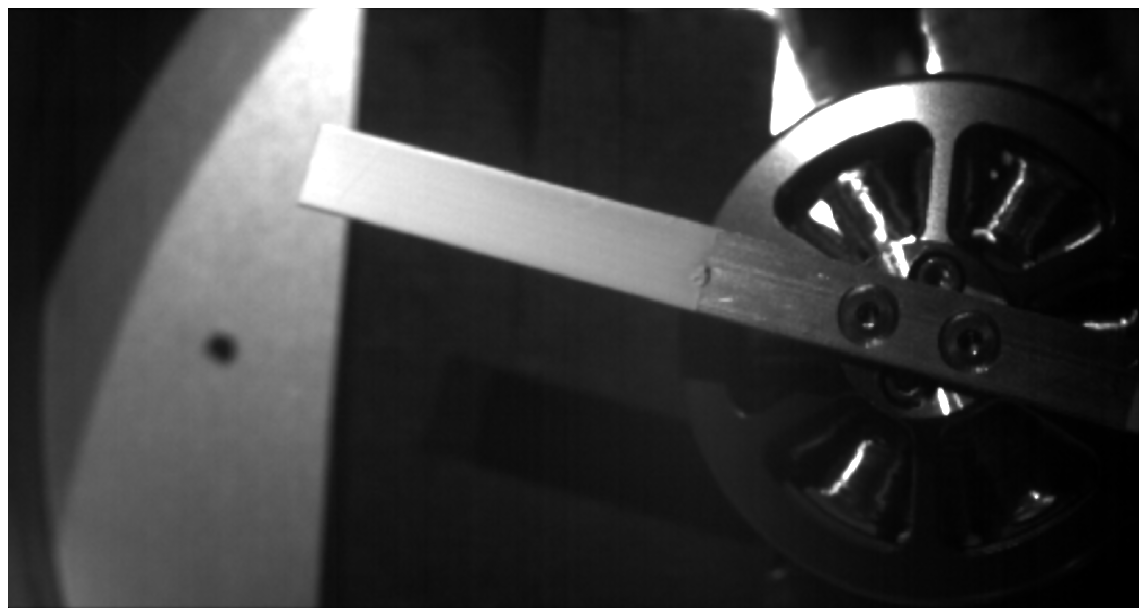}
        \subcaption{Image of static rotor}
        \label{fig:res_CW_stopped}
    \end{subfigure}
    \hfill
    \begin{subfigure}{0.4 \linewidth}
        \centering
        \includegraphics[trim = {20 70 20 0}, clip, scale = 0.38]{ 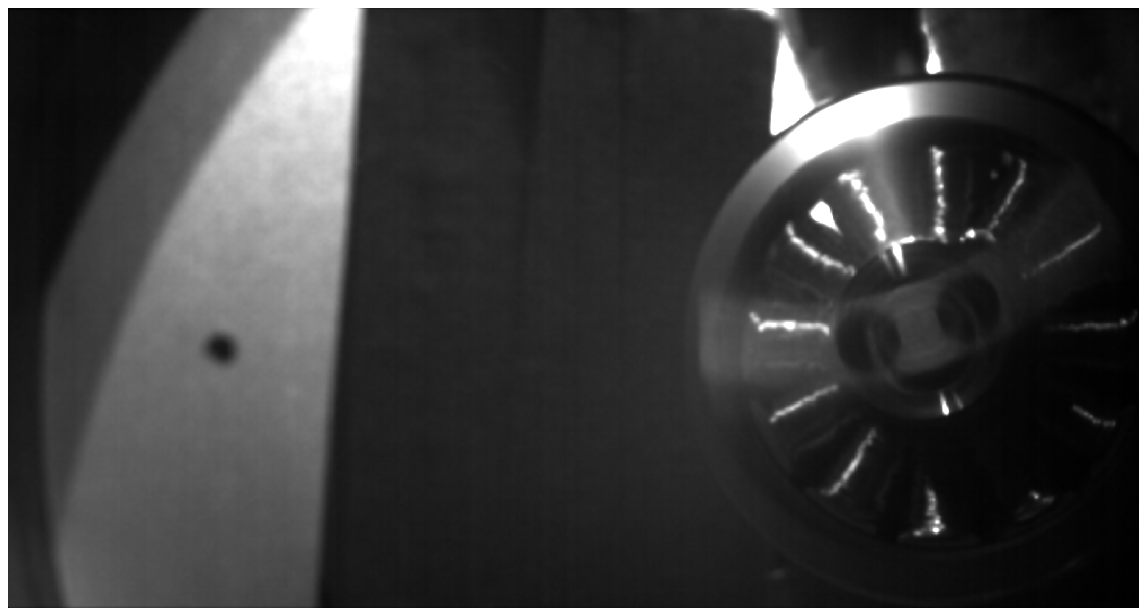}
        \subcaption{Image of moving rotor}
        \label{fig:res_CW_moving}
    \end{subfigure}
    \hspace{\fill}
    \caption[Motion Blur for continuous illumination]{Motion Blur for CW illumination with an exposure time less than \SI{1}{\milli\second}.}
\end{figure}

\begin{figure}[h]
    \hspace{\fill}
    \begin{subfigure}{0.4 \linewidth}
        \centering
        \includegraphics[scale = 0.35]{ 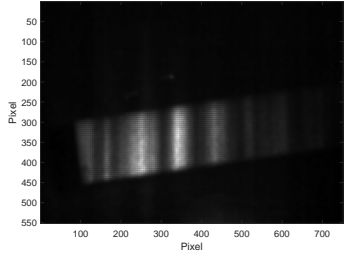}
        \subcaption{Image of static rotor}
        \label{fig:res_fs_stopped}
    \end{subfigure}
    \hfill
    \begin{subfigure}{0.4 \linewidth}
        \centering
        \includegraphics[scale = 0.35]{ 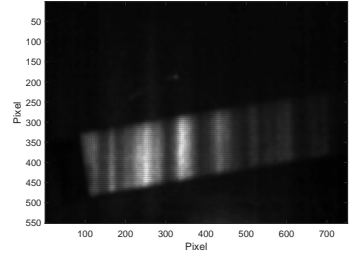}
        \subcaption{Image of moving rotor}
        \label{fig:res_fs_moving}
    \end{subfigure}
    \hspace{\fill}
    \caption[Absence of motion Blur for ultrashort illumination]{Motion Blur for ultrashort illumination with averaged intensity over all channels.}
    \label{fig:res_fs_img}
\end{figure}

The temporal resolution of the surface measurement can be tested by measuring the surface of a rapidly spinning rotor. A flat, \SI{20}{\centi\meter} long piece of aluminium is mounted on a drone motor and rotated at more than 25000 rpm. Based on an estimation of the motion blur, a temporal resolution of the measurement system in the range of \SI{100}{\nano\second} would be required, a value far below the temporal resolution of state of the art measurement setups. An image of a static and a moving rotor is shown in Fig.  \ref{fig:res_CW_stopped} and \ref{fig:res_CW_moving} respectively, which were recorded with a continuous illumination at \SI{1}{\milli\second} exposure time of the hyperspectral cameras. It is evident that the movement of the rotor cannot be resolved at this time resolution. Compared to this, Fig. \ref{fig:res_fs_stopped} and \ref{fig:res_fs_moving} show the ultrafast patterns projected onto the stopped and the moving rotor. The image of the moving rotor shows no recognizable motion blur. The exposure time of \SI{1}{\milli\second} ensures, that only one laser pulse is measured, due to the repetition rate of the laser amplifier of \SI{1}{\kilo\hertz}. A common trigger signal for both cameras ensures that the same pulse is measured.

\begin{figure}[h]
    \hspace{\fill}
    \begin{subfigure}[t]{0.4 \linewidth}
        \centering
        \includegraphics[scale = 0.36]{ 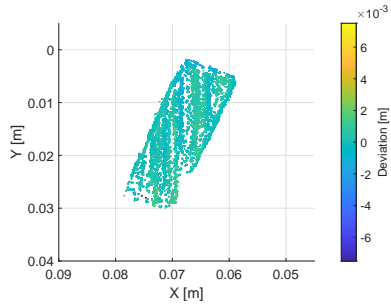}
        \subcaption{Point cloud of the calculated deviation from the fitted plane. The color scheme indicates the deviation.}
        \label{fig:res_diff_scatter_rotor}
    \end{subfigure}
    \hfill
    \begin{subfigure}[t]{0.4 \linewidth}
        \centering
        \includegraphics[scale = 0.36]{ 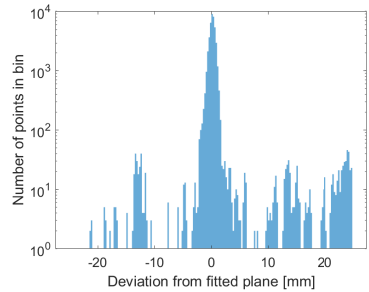}
        \subcaption{Logarithmic histogram with difference to plane fitted through the moving rotor surface}
        \label{fig:res_diff_hist_rotor}
    \end{subfigure}
    \hspace{\fill}
    \caption[3D reconstruction of a moving rotor]{3D reconstruction of moving rotor}
    \label{fig:res_diff_3d_rotor}
\end{figure}

Based on a measurement of the moving rotor, the surface data is reconstructed. The resulting point cloud is shown in Fig.  \ref{fig:res_diff_scatter_rotor}. As the cameras are not placed perpendicular to the rotor, a small gradient in the depth value is visible. Laterally, the size of the measurement volume is approximately \SI{4}{\centi\meter} by \SI{6}{\centi\meter}, which is the area where fringes are projected. Object space pixel resolution is approx. \SI{50}{\micro\meter} for individual subpixels. The 5-by-5 pixel thus has an object space resolution of \SI{250}{\micro\meter}. An interpolation is done to generate full resolution images for each color channel.  In z-direction the measurement volume is limited by the depth of focus of the camera objectives.

A plane surface is fitted through the measured point cloud, and the resulting difference to the reference plane is plotted in Fig. \ref{fig:res_diff_3d_rotor}. This histogram contains 200 bins in an interval of size \SI{2.5}{\centi\meter} on each side of the reference plane and is plotted logarithmically. Over four of these measurements, a mean standard deviation from the plane of \SI{2.5}{\milli\meter} and a mean percentage of reconstruction artifacts of \SI{1.95}{\percent} is measured. 

Due to the lack of motion blur, a temporal resolution of less than \SI{100}{\nano\second} can be demonstrated, which is a lower limit to the temporal resolution, as no measurements that require higher temporal resolution could be done for safety reasons. The estimated temporal resolution of the measurement setup is expected to be in the range of single picoseconds. This estimate is given by the output pulse duration, as seen in Fig.  \ref{fig:res_fiber_AC}. The dispersive effects of the diffractive pattern generation and the refractive projection is negligible for an input pulse with pulse duration \SI{1}{\pico\second}. 

These snapshot measurements with high temporal resolution can be done at the frame rate of the camera, which is given by the manufacturer as up to \SI{340}{\hertz}, while the light source supports a \SI{1}{\kilo\hertz} frame rate, given by the repetition rate of the laser amplifier. 

\begin{figure}[h]
    \hspace{\fill}
    \begin{subfigure}[t]{0.4 \linewidth}
        \centering
        \includegraphics[scale = 0.36]{ 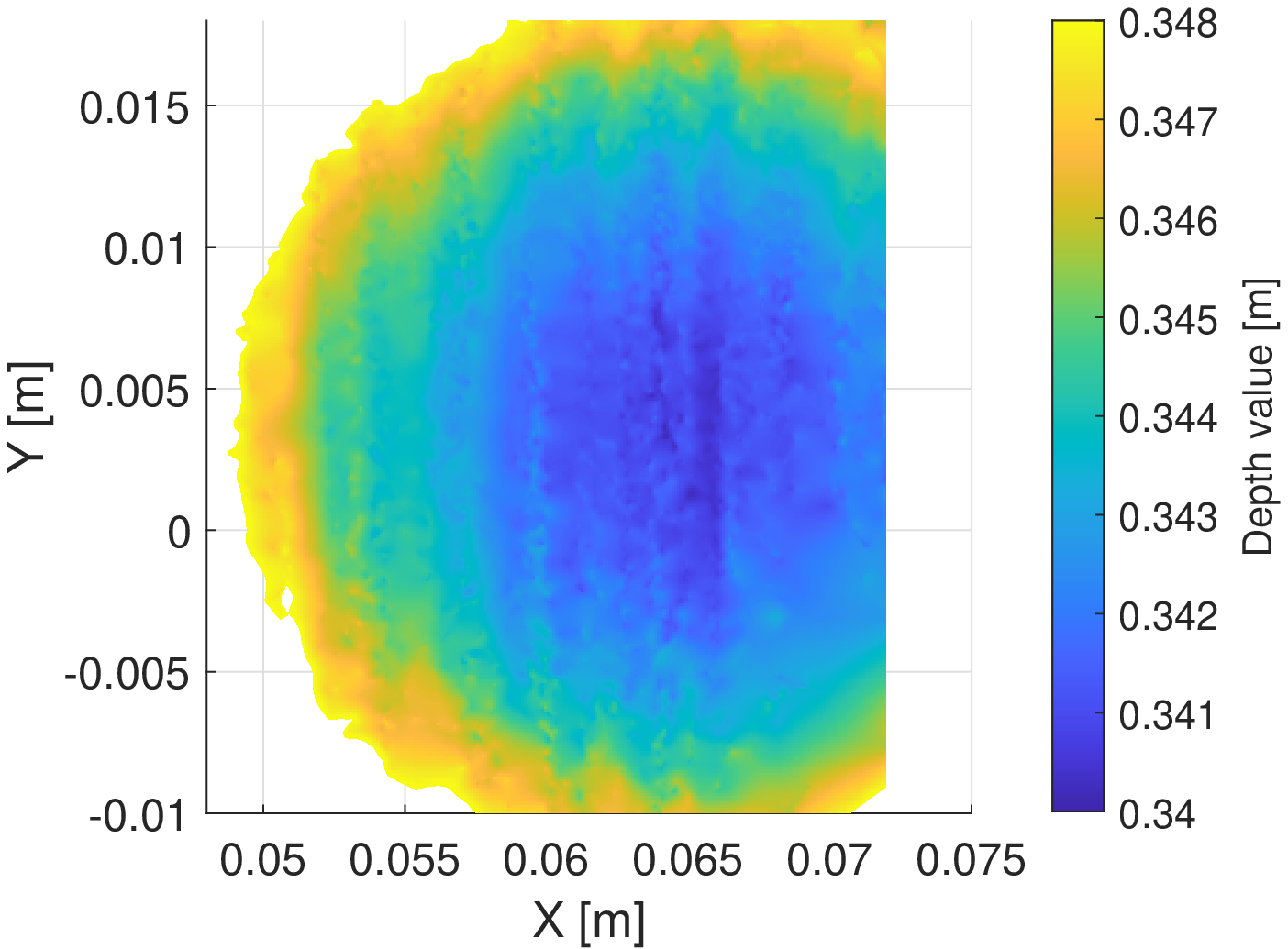}
        \subcaption{Reconstructed surface of the reference sphere based on measured depth values and Delaunay triangulation.}
        \label{fig:res_diff_scatter_sphere}
    \end{subfigure}
    \hfill
    \begin{subfigure}[t]{0.4 \linewidth}
        \centering
        \includegraphics[scale = 0.36]{ 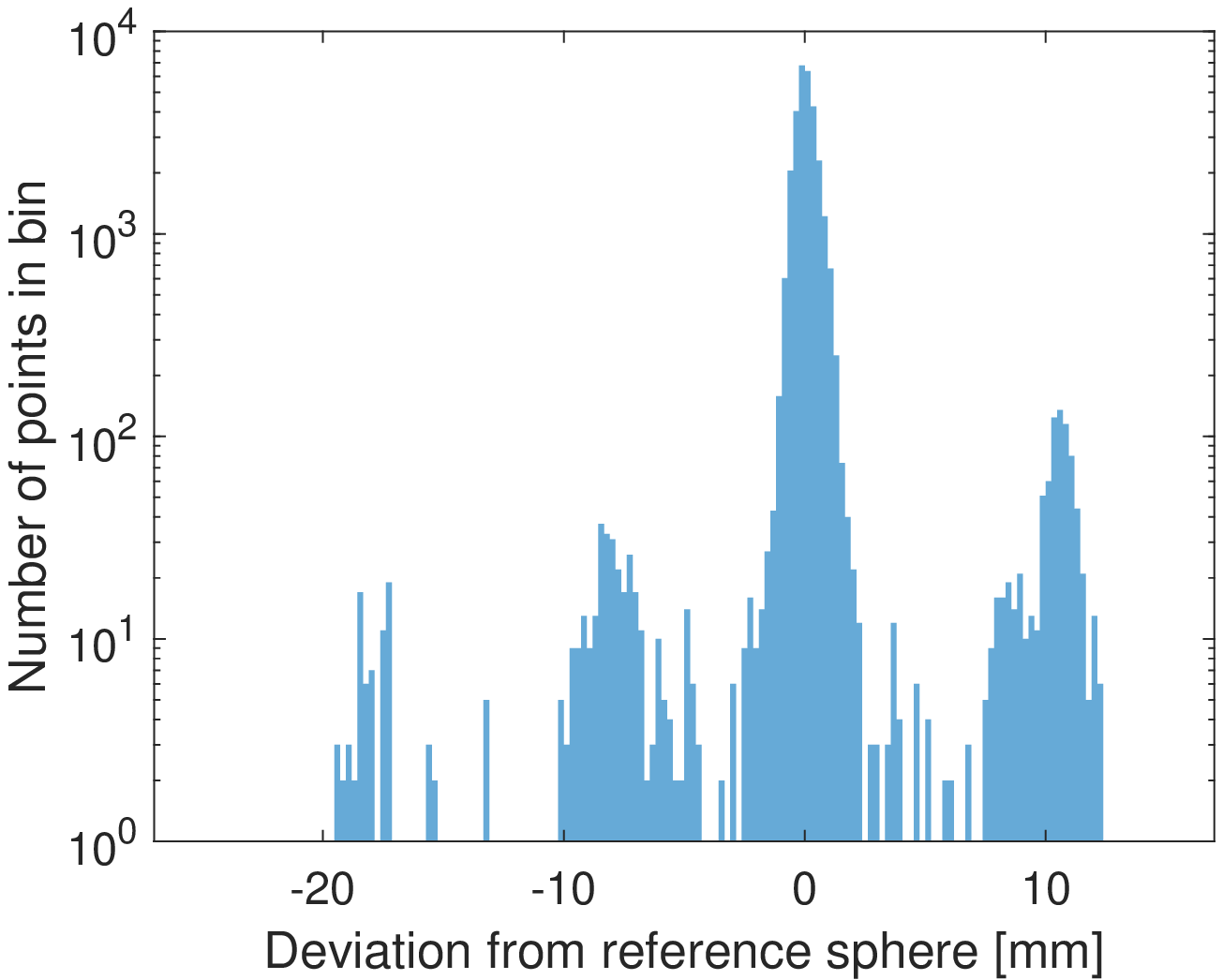}
        \subcaption{Logarithmic histogram with difference to the fitted sphere.}
        \label{fig:res_diff_hist_sphere}
    \end{subfigure}
    \hspace{\fill}
    \caption[3D reconstruction of static sphere]{3D reconstruction of a static sphere}
    \label{fig:res_diff_3d_sphere}
\end{figure}

 To investigate the measurement accuracy for more complex three dimensional structures, a static reference sphere with \SI{40}{\milli\meter} diameter was measured with the setup. For a homogeneous reflective behaviour, the sphere was coated with an opaque and diffusely reflecting paint. Based on the generated depth information for the illuminated spherical section, the surface was reconstructed with a Delaunay triangulation. The result is visible in Fig. \ref{fig:res_diff_scatter_sphere}. The difference of the measured depth values to an ideal sphere is calculated using a spherical fit through the data. Analogously to the plane measurement, this difference is plotted as a logarithmic histogram in Fig. \ref{fig:res_diff_hist_sphere}. The measurement performance is comparable to the plane measurement. 
 
\section{Conclusion and outlook}

The demonstrated approach enables the first 3D measurement of surfaces on ultrafast time scales. This is achieved by combining the well-known fringe pattern projection method for surface measurement with the spectral-spatial resolution of hyperspectral cameras as well as the high energy and ultrashort pulse duration of a femtosecond laser system, specifically enhanced by spectral broadening in a hollow core fiber. A system was realized, consisting of two synchronized hyperspectral cameras on CMOS basis with 25 spectral channels, fringe pattern generation utilizing high order diffraction and an amplified Ti:Sapphire laser, offering \SI{1}{\milli\joule} pulses with over \SI{100}{\nano\meter} bandwidth and a sub-ps pulse duration. 

By measuring a rotor at near sonic speeds a lower bound for the temporal resolution of the measurement system was established in the range of \SI{100}{\nano\second}. Despite the high speed of the measured object, no degradation due to motion blur was measured. The estimation for the maximum temporal resolution is given by the pulse duration of the broadened laser pulses, which is in the range of single picoseconds. The depth resolution was evaluated by the standard deviation of the measurement data to a fitted reference plane and was determined as \SI{2.5}{\milli\meter}. These measurements can be done at the frame rate of the hyperspectral cameras. 

Due to the extremely high temporal resolution, but comparatively low frame rate, the measurement system is well suited for measurements of reproducible processes, for example in conjunction with an optical delay stage. Possible areas of interest include laser induced plasma or surface deformation due to sound propagation in solid media.

This initial demonstration of the measurement principle opens up several possible avenues for optimization. With an increased spectral broadening during the fiber propagation, a more robust and accurate 3D reconstruction is expected due to the increased number of patterns. Similarly, complex diffractive elements can be engineered for the pattern generation, offering more design space in regards to the fringe pattern parameters, as well as the homogeneity of the illumination. Further optimization is possible in the hyperspectral camera technology, where a reduction in crosstalk as well as an increase in spectral channels could be achieved by using more sophisticated spectral filters.

\section*{Funding}

Bundesministerium für Bildung und Forschung (13XP5053A, 03ZZ0475, 13N15429)\\
Carl Zeiss Stiftung (Jena Alliance Life in Focus)

\section*{Acknowledgments}

The authors thank S. Börner of the Institute of Applied Physics for assistance with the vacuum setup and electronics. 
\\FE acknowledges support by the Zwanzig20 Research Alliance 3Dsensation.

\section*{Disclosures}

The authors declare no conflicts of interests.

\section*{Data availability} 

Data underlying the results presented in this paper are not publicly available at this time but may be obtained from the authors upon reasonable request.

\section*{Supplemental document}

\subsection*{Hyperspectral imaging}

Hyperspectral cameras can differentiate between a large number of spectral regions due to a Fabry-Perot filter matrix placed on the imaging sensor. Each Fabry-Perot etalon consist of two parallel reflective surfaces with seperation $h$ and reflectances $r_1, r_2$. The intensity transmittance is thus given by

\begin{equation}
    T = \frac{T_\mathrm{max}}{1 + \left( \frac{2F}{\pi} \right) ^2 \sin^2{\phi}},
\end{equation}
where $T_\mathrm{max}$ is 
\begin{equation}
    T_\mathrm{max} = \frac{(1-|r_1|^2)(1-|r_2|^2)}{(1-|r_1 r_2|)^2}
\end{equation}
and with the finesse $F$ defined as 
\begin{equation}
    F = \frac{\pi \sqrt{|r_1 r_2|}}{1 - |r_1 r_2|},
\end{equation}
as well as the phase $\phi$ given by the separation $h$, the refractive index $n$ between the interfaces, the angle of incidence 
\begin{equation}
    \phi = \frac{2 \pi n h \cos{\theta}}{\lambda}, \quad n \sin{\theta} = \sin{\theta_0}.
\end{equation}

\begin{figure}[hb]
  \centering
  \includegraphics[scale = 0.2]{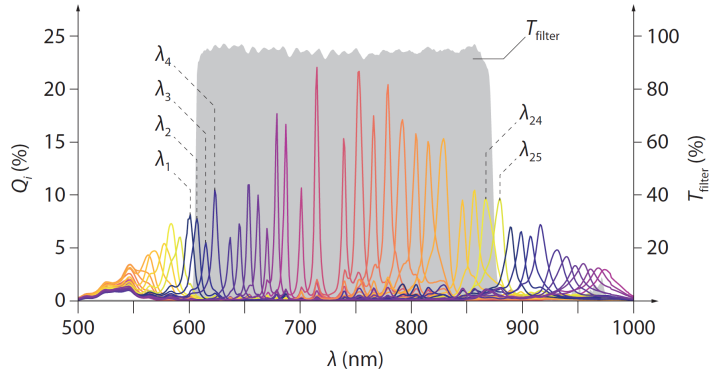}
    \caption{Quantum efficiency of the 25 different color channels generated by the 5-by-5 filter matrix of a hyperspectral camera (XIMEA MQ NIR). The grey area shows the transmittance of a bandpass filter for use of the camera in the red and near infrared  region. \cite{Heist.2018}}
    \label{fig:ximea_channels}
\end{figure}

Thus, the center wavelength of the transmission of a Fabry-Perot etalon is determined by the separation between the two surfaces. Notably, due to the width of the transmission spectra of the individual filters, a certain amount of overlap is present between the spectral channels. In Fig. \ref{fig:ximea_channels}, the spectral sensitivity of the camera is shown. This effect is also called crosstalk and is generally undesirable, as light from a different spectral region is attributed to the channel it is measured in. Other effects, like an imperfect placement of the filter matrix on the imaging sensor or leakage between adjacent pixels can also contribute to crosstalk. For the application in a spectrally multiplexed fringe projection system, this also leads to a reduce in contrast in the fringe patterns, and negatively affects the uniqueness of the individual patterns.

\printbibliography






\end{document}